\documentclass[10pt,letterpaper,twocolumn]{article}
\usepackage[english]{babel}
\usepackage{graphicx}

\newcommand{\alt}{\mathbin{\lower 3pt\hbox
   {$\rlap{\raise 5pt\hbox{$\char'074$}}\mathchar"7218$}}}
\newcommand{\agt}{\mathbin{\lower 3pt\hbox
   {$\rlap{\raise 5pt\hbox{$\char'076$}}\mathchar"7218$}}}

\begin{document}

\setcounter{footnote}{0}
\setcounter{equation}{0}
\setcounter{figure}{0}
\setcounter{table}{0}

\title{\large\bf How to observe the
localization law $\sigma(\omega)\propto -i\omega$ \\ for
conductivity?}

\author{\small  I. M. Suslov \\
\small Kapitza Institute for Physical Problems,
\\  \small Moscow, Russia \\{} \\
\parbox{120mm}{\footnotesize \,The Berezinskii localization
law $\sigma(\omega)\propto
-i\omega$ for frequency-dependent conductivity was never
questioned from the theoretical side,  but never observed experimentally. In
fact, this result is valid for closed systems, while  most of actual systems
are open. We discuss several possibilities for observation of this law and
 experimental difficulties arising at this way. } }

\date{}

\maketitle

It is well-known \cite{100,101} that the electron states in
disordered systems can be extended or localized. In the latter
case, when a system is an Anderson dielectric, its
frequency-dependent conductivity is believed to obey  the Mott
law, ${\rm Re}\, \sigma(\omega)\propto \omega^2
\ln^\alpha{\omega}$. In fact, in the low-frequency limit
conductivity is dominated by its imaginary part, and dependence
$\sigma(\omega) \propto\,-i\omega$  is expected.

The  localization law $\sigma(\omega)\propto\,-i\omega$
was predicted by
Berezinskii in 1973 \cite{1} for one-dimensional disordered
systems. According to self-consistent theory
 by Vollhardt and W$\ddot o$lfle \cite{2}, the same result
is valid in the localization phase for systems of
arbitrary dimension $d$. In the recent paper  \cite{3} of the
present author the same behavior of conductivity was established
for systems of finite size $L$ at the arbitrary extent of
disorder.  The latter is a consequence of the fact that a finite
system is topologically zero-dimensional, and  its effective
dimensionality is less than lower critical one ($d_{c1}=2$
\cite{102}).

The Berezinskii law was never questioned in the theoretical
community; however, it was never observed
experimentally. This paradoxical situation  was clarified
in Ref.5:  Berezinskii's result is valid in
closed systems, while  most of actual systems are open. In
open systems, replacement  $ -i\omega \to -i\omega +\gamma$
occurs (where $\gamma$ is inelastic damping) and
dependence $\sigma(\omega)\propto -i\omega$ transforms into
the usual metallic behavior.

A possibility of realization of closed systems
became clear after observation of the persistent current
in disordered systems (in the Aharonov--Bohm geometry)
 \cite{4,5,6},
in accordance with its prediction in  \cite{7}. In
 fact, the persistent current is a consequence of the
 Berezinskii law,  establishing the dissipativeless character of
 conductance.  Its  observation is possible, when  a size  $L$ of
 the disordered ring is  small in comparison with the inelastic
 length  $L_{in}$, depending on temperature $T$.  The typical
scales in the indicated experiments were $L\sim 1 \mu m$, $T\sim
100 mK$.  If one accepts that $L_{in}\propto T^{-2}$ (as for
e-e interaction), then a system is closed for $L\alt  10 nm$ in
the helium region ($T\sim 1K$).

Let discuss several experimental situations, where observation of
the Berezinskii law is possible.   \vspace*{2mm}

1. The first variant is the island film of a disordered metal
lying on the dielectric substrate (Fig.1). We suppose for
clearance that all islands are of  the same size $L$, which
increases monotonically in the course of the film
deposition\,\footnote{\,In fact, there is a distribution of
islands in size, which shifts to the large $L$ region in the
course of deposition. }.  Then for $L\alt L_{in}$ the Berezinskii
law is valid (Fig.1,a), while in the opposite case $L\agt L_{in}$
the usual metallic conductivity takes place (Fig.1,b). A
transition from one regime to another can be provided by the
change of  $L$ or the temperature.

\begin{figure*}
\centerline{\includegraphics[width=5.1 in]{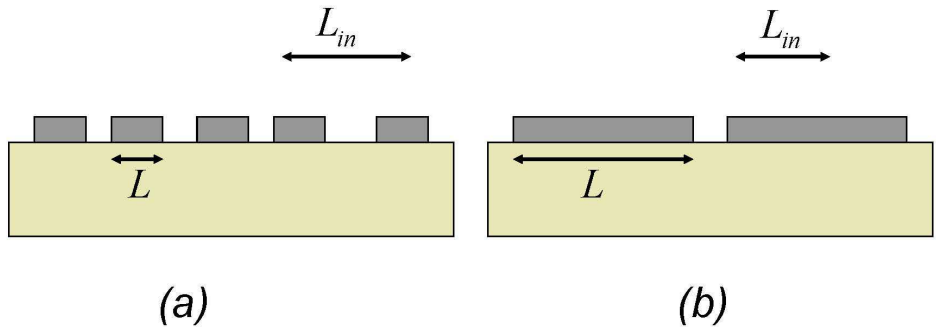}}
\caption{\small In the case of the island metallic film, the
 Berezinskii law
is observable when the island size  $L$ is small in comparison
with the inelastic length $L_{in}$ (a), while in the
opposite case the metallic behavior is valid (b).  }
\label{fig1} \end{figure*}

At first glance, the described experiment is simple. However,
there is a bottleneck in it. It is clear from  relation
$\epsilon\sim i\sigma/\omega$, that behavior
$\sigma\propto -i\omega$ corresponds to the
frequency-independent  permittivity
%dielectric permeability
$\epsilon$, so a disordered system is an ordinary
 dielectric. The properties of the film in
 the Berezinskii law regime are the same as those of the
dielectric substrate, hence the former gives a negligible
contribution to conductivity in comparison with the latter.
The width of the film is by 6--7 orders less than the
width of the substrate, but the corresponding smallness can be
partially compensated by a large value of the film permittivity
$\epsilon_1$  in comparison with its substrate value
$\epsilon_0$. By the order of magnitude,
$\epsilon_1\sim \xi^2/a_0^2$ (where $\xi$ is the localization
length for wave functions, and  $a_0$ is the atomic space),
and saturates by a value  $L^2/a_0^2$ for large  $\xi$.
If the metallic film is weakly disordered\,\footnote{\,The
films are weakly disordered in the case of "simple" metals
(such as Mg, Al, Sn), which are well-described by the
pseudopotential theory \cite{8}; a small pseudopotential
 provides weak scattering even in the amorphous state.
Contrary, the films of the transition metals are usually strongly
disordered. }, then for
$L\sim 10 nm$ its permittivity $\epsilon_1$ can exceed
$\epsilon_0$  by  3--4 orders.

The experimental procedure looks as follows. The
experiment is carried out  $in$ $citu$ and begins with a
measurement of the frequency and temperature dependencies
of the substrate conductivity, with saving results in a
file. Then a small amount of metallic atoms is deposited, and
again conductivity is measured and saved; again deposition is
made an so on. Proceeding by small steps, one should reach a
regime, when  the film contribution is clearly seen in
the substrate background. Then the actual measurements
can be made.

\vspace*{2mm}
\begin{figure}
\centerline{\includegraphics[width=1.9 in]{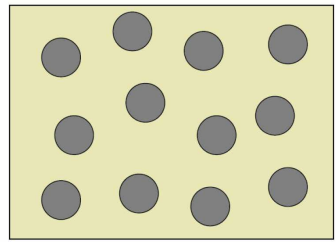}}
\caption{\small  A nanocomposite system with spherical
 metallic granules
embedded in a dielectric.  } \label{fig2} \end{figure}

2. The second example is the nanocomposite system  \cite{9,9a},
which is a dielectric sample with the metallic granules embedded
in it (Fig.2). The volume fraction  $p$ of a metal can be rather
large and its effect should be easily observable, being  of the
order of unity. However, a "hidden rock" is present here. Let
exploit the formula from the Landau and Lifshitz book \cite{11}
$$
\frac{\bar \epsilon -\epsilon_0}{\epsilon_0 } =
p\, \frac{3( \epsilon_1 -\epsilon_0)}{2\epsilon_0 +\epsilon_1}
 \,,
\eqno(1)
$$
which is valid for a small concentration of spherical granules: it
gives the average permittivity $\bar \epsilon$ (for the system of
Fig.2) in terms of its values for a dielectric ($\epsilon_0$)
and a metal ($\epsilon_1$).  Since $\epsilon_1\gg \epsilon_0$,
then
$$
\frac{\bar \epsilon -\epsilon_0}{\epsilon_0 } \approx
3p -9 p \frac{\epsilon_0}{\epsilon_1} \,,
\eqno(2)
$$
and the main
contribution $3p$ is an uninteresting constant, while the useful
effect, depending on  $\epsilon_1$, is determining by two small
parameters  $p$ and $\epsilon_0/\epsilon_1$. As a result, the
problem of a reference arises, i.\,e. a necessity to have the
identical sample  without metallic granules. Fortunately, such a
problem is absent for a specific technology \cite{9,9a,10}, when
nanocomposites are produced on the base of a porous glass, whose
pores are filled by metallic granules (of suitable size about
$7nm$); so the same sample can be measured in absence and in
presence of granules. It is useful to note that for the system of
Fig.2 (in contrast to that of Fig.1) the strongly disordered
metal is desirable, in order to increase the ratio
$\epsilon_0/\epsilon_1$.

\vspace*{2mm}

3. Derivation of Eq.1 is based on a solution of the well-known
problem on a dielectric ball in the external electric field
\cite{11}. The analogous problem is solvable for an ellipsoid
with arbitrary ratios of its semi-axes $a$, $b$, $c$ \cite{11},
and generalization of (1) is possible for granules of
ellipsoidal form:
$$
\frac{\bar \epsilon -\epsilon_0}{\epsilon_0 } =
p\, \frac{ \epsilon_1 -\epsilon_0}{A\epsilon_0 + B \epsilon_1}
\,, \eqno(3)
$$
where    $A=1-B$, and
$$
B = \frac{abc}{2} \, \int_0^\infty \,
\frac{dx} {(x+a^2)^{3/2} (x+b^2)^{1/2} (x+c^2)^{1/2}} \,,
\eqno(4)
$$
if the electric field $\bf E$ is  directed along
the axis $a$.

In reality, the metallic granules are not strictly spherical. For
modelling of such situation, one can suggest that granules are
ellipsoids with fluctuating ratios of semi-axes. Then for
$\epsilon_1\gg \epsilon_0$ one has
$$
\frac{\bar \epsilon -\epsilon_0}{\epsilon_0 } \approx
 p \left\langle B^{-1} \right\rangle -
p \left\langle B^{-2} \right\rangle
\frac{\epsilon_0}{\epsilon_1}
\eqno(5)
$$
($\langle \ldots \rangle$ is averaging over fluctuations), so
the structure of Eq.2 is preserved but the coefficients are
changed.

Parameter  $B$ decreases when $a$ becomes greater than
$b$ and $c$. In the limit of a strongly oblong ellipsoid
($a\gg b\sim c$) one has  $B\to 0$ and Eq.3 takes a form
$$
\frac{\bar \epsilon -\epsilon_0}{\epsilon_0 } =
p\, \frac{ \epsilon_1 -\epsilon_0}{\epsilon_0 } \,,
\eqno(6)
$$
i.\,e. optimal conditions for observation correspond to the
needle-shaped granules  (Fig.3). In this case one can provide a
sufficient smallness of $p$ (which is necessary for validity of
Eq.3 and a transparent interpretation of the experiment) and its
compensation by a large parameter $\epsilon_1/\epsilon_0$. As a
result, the effect is of the order of unity, or even more. Such
systems can be fabricated on the base of chrysotile asbestos,
which is a stack of the parallel nanotubes \cite{12} with a
typical pore diameter $5nm$; since the length of the granules
should be essentially greater\,\footnote{\,This length can reach
$1mm$ \cite{12a}.}, one is induced to work in the millikelvin
range of temperatures.

\begin{figure}
\centerline{\includegraphics[width=2.1 in]{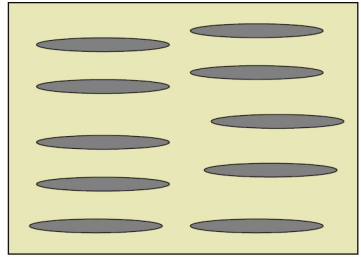}}
\caption{\small  A nanocomposite system with
 needle-shaped  granules.}
\label{fig3} \end{figure}

\vspace*{5mm}

4. In relation with the latter,
we can
indicate one
exotic possibility. If a vessel with superfluid helium is
rotated, then a set of the parallel vortices arises. If
metallic atoms are injected in helium, they are localized at
the vortex cores and form nanowires \cite{13}. Regulating the
length of the latter, one can create the desired system
(Fig.4). A concentration of the metallic phase is strongly
restricted in this case\,\footnote{\,For realistic conditions,
one can have about $10^4$ vortices per $1cm^2$, while a
diameter of nanowires varies from $a_0$ till several
nanometers.}, but at sufficiently low temperatures one can deal
with large $L$ scales and, as a consequence, with enormous values
of permeability $\epsilon_1$.  \vspace*{2mm}

Analogously, parameter  $B$
tends to zero  in the case of pancake-shaped granules  ($a\sim b
\gg c$), if their plane is oriented along the electric field;
this case is also described by formula (6). In particular, it is
valid in the situation of Fig.1, where the volume concentration
$p$ is inevitably small.
%\vspace*{2mm}

In conclusion, the Berezinskii law was not observed previously,
since it refers to closed systems, while most of actual systems
are open. We have suggested several possibilities for its
observation, and shown that experimental difficulties are present
in all situations. The latter is rather natural, since in the
opposite case this law would be discovered experimentally long
ago.

\vspace*{2mm}
\begin{figure}
\centerline{\includegraphics[width=2.5 in]{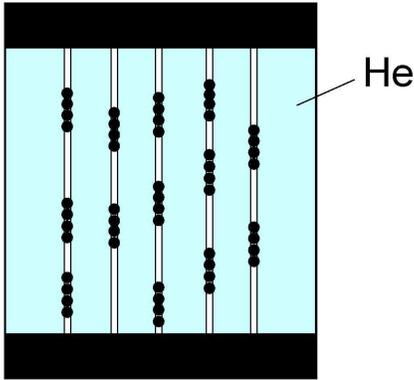}}
\caption{\small  An
exotic realization of the system represented in Fig.3. If a vessel
with superfluid helium is rotated, then a set of the parallel
vortices arises, and the injected metallic atoms are localized on
the vortex cores.
  }
\label{fig4} \end{figure}

The author is indebted to S.~V.~Demishev, E.~Yu.~Koroleva,
P.~I.~Arseev, E.~B.~Gordon, F.~A.~Pudonin for useful discussions.
%\vspace*{2mm}

%\newpage

\end{document}